\documentclass[aps,prb,twocolumn,superscriptaddress]{revtex4}
\usepackage{amsfonts}
\usepackage{amsmath}
\usepackage{amssymb}
\usepackage{graphicx}

\begin{document}

\title{Effect of impurity resonance states on the NMR spectra of high-$T_c$ cuprates}

\author{J. Chang}
\affiliation{Institute of Theoretical Physics and
Interdisciplinary Center of Theoretic Studies, Chinese Academy of
Sciences, P.O. Box 2735, Beijing 100080, China}
\author{Y. H. Su}
\affiliation{Center for Advanced Study, Tsinghua University,
Beijing 100084, China}
\author{H. G. Luo}
\affiliation{Institute of Theoretical Physics and
Interdisciplinary Center of Theoretic Studies, Chinese Academy of
Sciences, P.O. Box 2735, Beijing 100080, China}
\author{H. T. Lu}
\affiliation{Institute of Theoretical Physics and
Interdisciplinary Center of Theoretic Studies, Chinese Academy of
Sciences, P.O. Box 2735, Beijing 100080, China}
\affiliation{Department of Physics, Peking University, Beijing
100084, China}
\author{T. Xiang}
\affiliation{Institute of Theoretical Physics and
Interdisciplinary Center of Theoretic Studies, Chinese Academy of
Sciences, P.O. Box 2735, Beijing 100080, China}
\affiliation{Center for Advanced Study, Tsinghua University,
Beijing 100084, China}

\begin{abstract}
A strong nonmagnetic impurity can induce a resonance state in the $d$-wave
superconducting state. As far as magnetic properties are concerned, this resonance state
behaves effectively like a free moment. It leads to a Curie-Weiss-like magnetic
susceptibility in an intermediate temperature regime below $T_{c}$. From the impurity
susceptibility, the effective moment of the resonance state is deduced and compared with
experiments. The contribution of the resonance to the magnetic susceptibility can account
for the main feature of the NMR spectra in overdoped high-$T_c$ materials. In the
underdoped regime, the contribution from the resonance to the magnetic susceptibility is
also substantial, but the effective moment of the resonance is smaller than the total
moment induced by a nonmagnetic impurity.

\end{abstract}

\pacs{74.25.Jb,76.60.-k}
\maketitle

The substitution of impurities into cuprates has served as an important probe in the
study of high-temperature superconductivity. Experiments were performed with magnetic
(Ni) as well as nonmagnetic (Zn, Li, Al) impurities. Ni$^{2+}$ carries a local moment. It
leads to naturally some kind of Kondo physics. However, the behavior of nonmagnetic
impurities is surprising. Theoretical analysis indicates that Zn or some other
nonmagnetic impurities in CuO$_{2}$ planes is a strong potential scatterer. It can induce
a sharp resonance peak near the Fermi level in a $d$-wave superconducting state.
\cite{Balatsky,Salkola} This low-energy resonance peak was observed in the scanning
tunneling spectroscopy (STM) measurement around a Zn impurity \cite{S.H.Pan} and the STM
pattern associated can be understood by considering the effect of tunneling
filter.\cite{Martin,Xiang}

However, the interpretation to the experimental data obtained from magnetic measurements
is still controversial.\cite{Alloul,Mahajan,Julien,Bobroff,Wang,G.-M. Zhang,XiangW} A
series of nuclear magnetic resonance (NMR) experiments, such as the $^{89}$Y NMR width of
YBa$_2$(Cu$_{0.96}$Zn$_{0.04}$)$_3$O$_{6+x}$ samples with $x \geq 0.5 $ showed that the
magnetic susceptibility associated with Zn or other nonmagnetic impurities exhibits a
Curie-like behavior. \cite{Alloul,Mendels,Mahajan,Julien,Ishida,Bobroff} This led to the
suggestion that some uncompensated spins are induced by Zn in a correlated background
with strongly antiferromagnetic fluctuations. However, muon spin resonance ($\mu$SR)
measurements on a number of YBa$_2$(Cu$_{1-x}$Zn$_x$)O$_8$ samples with different Zn
content found no evidence for an additional local moment induced by Zn above the
background.\cite{Bernhard} It was suggested that the Curie-like Knight shift observed in
NMR experiments may result from a sharp low-energy peak in the density of states already
observed by STM, rather than from induced local moments.\cite{Williams} In the overdoped
regime, since the magnetic correlation is dramatically suppressed, the formation of
induced local moments has also been questioned.\cite{Tallon}

In this paper, we investigate the effect of resonance states induced by a strong
nonmagnetic impurity on NMR spectra in a $d$-wave superconducting state. The contribution
of the resonance state to the spin-lattice relaxation rate as well as the Knight shift is
shown to be Curie-Weiss-like. From the impurity susceptibility, the effective moment of
the resonance state is determined and compared with experiments. Our result is obtained
in the superconducting state. We believe that it can be also qualitatively applied to the
pseudogap phase of high-T$_c$ oxides, since the normal state pseudogap has similar
symmetry as the $d$-wave superconducting gap.

Let us consider the following BCS mean-field Hamiltonian of $d$-wave superconductors
with a nonmagnetic impurity.
\begin{equation}  \label{Hamiltonian}
H=\sum_{k}C_{k}^{\dagger }(\xi _{k}\tau _{3}-\Delta _{k}\tau _{1})C_{k}+
\frac{V}{N}\sum_{k,k^{\prime }}C_{k}^{\dagger }\tau _{3}C_{k^{\prime }} ,
\end{equation}
where $C_{k}^{\dagger }=(c_{k\uparrow }^{\dagger }, c_{-k\downarrow })$ and $ \xi_k$ is
the energy dispersion of normal electrons. $\tau _{1}$ and $\tau _{3}$ are the Pauli
matrices. $\Delta_k =\Delta (T)\cos (2\varphi_k )$ is the temperature-dependent gap
function of $d$-wave superconductors and $\varphi_k=\tan^{-1} k_y/k_x$. V is the impurity
scattering potential at $r=0$.

For the $\delta$-function scattering potential defined in Eq. (\ref{Hamiltonian}), only
the $s$-wave scattering channel has a contribution to the scattering matrix. In this
case, the Green's function of electron can be rigorously expressed as \cite{Balatsky,
Salkola, Hirschfeld}
\begin{equation}  \label{Green}
G(r,r^{\prime },\omega ) = G^{0}(r-r^{\prime },\omega )+ G^{0}(r,\omega
)T(\omega )G^{0}(-r^{\prime },\omega ),
\end{equation}
where $G^{0}(r,\omega )$ is the unperturbed propagator, which can
be obtained from the Fourier transformation of the Green's
function $ G^0(k,\omega)$ in the momentum space
\begin{equation*}
G^{0}(k,\omega) = \frac1{\omega - \xi_k\tau_3 -\Delta_k\tau_1}.
\end{equation*}
For a $d$-wave superconductor, the $T$ matrix $T(\omega )$ is diagonal and determined by
the following equations
\begin{eqnarray}
T(\omega )&= & T_{0}(\omega )\tau _{0}+T_{3}(\omega )\tau _{3} , \nonumber\\
T_{0} & =& \frac1{\pi N_F} \frac{G_{0}(\omega )}{c^{2}-G_{0}(\omega )^{2}},\nonumber\\
T_{3}&=& \frac1{\pi N_F}\frac{c}{c^{2}-G_{0}(\omega )^{2}},
\end{eqnarray}
where $N_{F}$ is the normal density of states at the Fermi level
and
\begin{equation*}
G_{0}(\omega )=\frac{1}{\pi N_{F}}\sum_{k}\frac{\omega}{\omega^2 - \xi^2_k
-\Delta_k^2}.
\end{equation*}
$c=\cot \delta _{0} = 1/(\pi N_F V)$ and $\delta _{0}$ is the
scattering phase shift. In the strong scattering (or unitary)
limit $\delta_0 \sim \pi/2 $ and $c \sim 0$. In the weak
scattering (Born) limit, $\delta_0 \rightarrow 0$ and $c
\rightarrow \infty$. In the discussion below, only the unitary
scattering limit will be considered.

In the complex $\omega$ plane, $T(\omega )$ has two poles when the
condition $ G_0(\omega ) = \pm c$ is satisfied. These two poles
define the resonant states induced by an impurity and are present
at
\begin{equation*}
\Omega_\pm =\Omega ^{\prime } \pm i\Omega ^{\prime \prime },
\end{equation*}
where $\Omega ^{\prime }$ denotes the resonance energy and $\Omega
^{\prime \prime }$ the resonance peak width. For a Zn impurity,
$\delta _{0}\approx 0.48\pi $ and $c\approx 0.0629$, both $\Omega
^{\prime }$ and $\Omega ^{\prime\prime }$ are found to be much
smaller than the superconducting gap, $(\Omega ^{\prime },\Omega
^{\prime \prime })\ll \Delta $. Thus a sharp resonance state is
induced by a Zn impurity.

In the low-energy limit, the unperturbed Green's function $G^0(r, \omega)$ changes
smoothly with $\omega$ and $\text{Im} G^0(r, \omega)$ approaches zero in the limit
$\omega \rightarrow 0$. This means that the imaginary part of $G^{0}(r,\Omega ^{\prime })
$ at the resonance energy is much smaller than its real counterpart and can be
approximately taken as zero once $r\not= 0$. In this case, the impurity correction to the
Green's function is then approximately given by
\begin{equation}
\delta G(r,r^\prime \omega ) \approx \text{Re}G^{0}(r,0)T(\omega )\text{Re}
G^{0}(-r^{\prime },0).  \label{correction}
\end{equation}

The NMR spin-lattice relaxation and Knight shift are proportional to the imaginary and
real parts of the magnetic susceptibility, respectively. To study the impurity
corrections to these quantities, we use the Hamiltonian first proposed by Mila and Rice
for the hyperfine interaction between nuclear spins and conduction electrons. This
Hamiltonian contains direct hyperfine interactions as well as exchange-induced hyperfine
interactions between neighboring sites. The spin-lattice relaxation rate at site $r$ is
then given by \cite{Millas,Moria}
\begin{equation}
\frac{1}{T_{1}(r)T}=\frac{k_{B}}{4\mu _{B}^{2}\hbar ^{2}}
\sum_{j,l}F_{j,r}F_{l,r}\lim_{\omega \rightarrow 0} \frac{\chi ^{\prime
\prime }(j,l,\omega )}{\omega }  \label{T1T}
\end{equation}
where $j$ or $l$ runs over $r$ and its four nearest neighbors.
When $j=r$, $F_{j,r}=A$ is the direct hyperfine coupling constant
between the nuclei and the electrons on the same site. When
$j\not= r$, $F_{j,r}=B$ is the hyperfine coupling induced by the
exchange interaction of Cu spins on the two nearest neighboring
sites.

In the limit $\omega \rightarrow 0$, the magnetic susceptibility $\chi
^{\prime \prime }(j,l,\omega )$ is given by
\begin{equation}
\lim_{\omega \rightarrow 0}\frac{\chi ^{\prime \prime }
(j,l,\omega )}{ \omega }=\frac{\mu _{B}^{2}\beta }{4\pi
}\int_{-\infty }^{\infty }d\varepsilon \frac{A(j,l, \varepsilon
)}{\cosh ^{2}(\beta \varepsilon /2)}, \label{x''}
\end{equation}
where
\begin{equation*}
A(j,j^{\prime },\varepsilon )=[\mathrm{Im}G_{11}(j,j^{\prime };
\varepsilon )]^{2}+[\mathrm{Im}G_{12}(j,j^{\prime };\varepsilon
)]^{2},
\end{equation*}
and $\beta =1/k_{B}T$. In the clean limit, $V=0$, $A(j,j^\prime ,\varepsilon )\sim
\varepsilon ^{2}$ in the low-energy limit $\varepsilon \ll \Delta $. It can be readily
shown from Eq. (\ref{x''}) that $1/T_{1}\sim T^{3}$. This $ T^{3}$ behavior of $1/T_{1}$
was observed in high-$T_{c}$ cuprates, in support of d-wave superconductivity.

At low temperatures, the spin-lattice relaxation rate is mainly determined by the
resonance state. Using Eq. (\ref{correction}) and assuming particle-hole symmetry, it is
straightforward to show that the contribution of the resonance state to the spin-lattice
relaxation rate is approximately given by
\begin{equation}
\delta \left[ T_{1}(r)T\right] ^{-1}\simeq \frac{k_{B}}{2\pi \hbar ^{2}}
Z^{2}(r)\int_{-\infty }^{\infty }d\varepsilon \left[ T_{11}^{\prime \prime }(\varepsilon
)\right] ^{2}P(\varepsilon ,T) , \label{dT1T}
\end{equation}
where
\begin{eqnarray*}
&& Z(r) = \sum_{j}F_{j,r}\{ \left[ \text{Re}G_{11}^{0}(j,0)\right] ^{2}+
\left[ \text{Re}G_{12}^{0}(j,0)\right] ^{2}\} , \\
&& P(\varepsilon ,T) =\frac{1}{4k_{B}T\cosh ^{2}(\varepsilon
/2k_{B}T)}.
\end{eqnarray*}
In the intermediate-temperature regime, $k_{B}T_{c}\gg k_{B}T\gg \Omega ^{\prime }$, the
integration in Eq. (\ref{dT1T}) is contributed mainly from the pole of
$T_{11}(\varepsilon )$, thus one can replace approximately $ P(\varepsilon ,T)$ by
$P(\Omega ^{\prime },T)$. In this case, $P(\Omega ^{\prime },T)\sim 1/T$, and
\begin{equation*}
\delta \left[ T_{1}(r)T\right] ^{-1}\sim \frac{1}{T}.
\end{equation*}
This $1/T$ behavior of $(T_{1}T)^{-1}$ was observed in Zn-substituted $
\mathrm{YBa_{2}Cu_{4}O_{8}}$ samples in the superconducting state. \cite {WilliamsNQR} It
is also consistent with the $^{63}\text{Cu}$ NMR data for Zn-substituted
$\mathrm{YBa_{2}Cu_{3}O_{6.7}}$.\cite{Julien}

In the low-temperature limit, $k_BT \ll \Omega^\prime$, since $P(T)$ drops to zero
exponentially with decreasing temperature, the impurity correction to the spin-lattice
relaxation rate is exponentially small and negligible. In the NMR experiments, the fast
drop of $(T_1T)^{-1}$ was generally taken as an indication of spin freezing.
\cite{Julien} However, the drop here is due to the fact that the resonance state has a
finite excitation energy above the Fermi level and is difficult to be excited when
$k_{B}T\ll \Omega ^{\prime }$.

Thus $\delta [T_1(r)T]^{-1}$ varies nonmonotonically with temperature. It first increases
with decreasing temperature and then drops after reaching a maximum. In the unitary
scattering limit, the peak temperature $T_{f}$ of $ \delta [T_1(r)T]^{-1}$ is
approximately given by $k_{B}T_{f}\simeq 0.65\Omega ^{\prime }$. For Zn-substituted
materials, the induced resonance state energy $\Omega ^{\prime } \sim 17$ K.
\cite{S.H.Pan} The corresponding peak temperature is estimated to be $T_f \sim 11$ K, in
agreement with the experimental data for
$\mathrm{YBa_{2}Cu_{3}O_{6.7}}$(YBCO),\cite{Julien} where the peak of $(T_1T)^{-1}$ is
located at $\sim 10$ K.

\begin{figure}[tb]
\includegraphics[width=6cm]{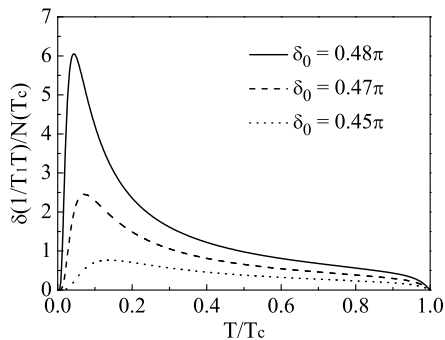}
\vspace{-0.2cm} \caption{Temperature dependence of the impurity
contribution to the spin-lattice relaxation rate on one of the
nearest neighboring sites of the impurity, normalized by the total
spin-lattice relaxation rate $N(T)= (T_1T)^{-1}$ at $T_c$. The
value of the phase shift is given in the figure. \label{fig1}}
\end{figure}

Figure \ref{fig1} shows the temperature dependence of $\delta (1/ T_{1}T)$, normalized by
the total spin-lattice relaxation rate $N(T)=(T_{1}T)^{-1}$ at $T_c$, on one of the four
nearest neighbors of the impurity. In obtaining the curves shown in this figure, the
energy dispersion $\varepsilon_{k}$ defined in Ref. \onlinecite{Norman} and the
zero-temperature superconducting gap $\Delta$(0 K)=25 meV are used. In this case, the
normal state density of states $N_F$ is about 1.9 eV$^{-1}$. The results show that the
impurity correction to the spin-lattice relaxation rate is very sensitive to the value of
the phase shift $\delta_0$. $\delta (1/T_1T)$ drops quickly with decreasing $\delta_0$.
In the limit $\delta_0 = \pi /2$, $\delta (T_1T)^{-1}$ increases monotonically with
decreasing temperature.

The Knight shift is another important quantity measured by NMR experiments. It is
determined by the real part of the static magnetic susceptibility $ \chi ^{\prime }$
(Ref.\onlinecite{Bulut}),
\begin{equation}
K(r)=\frac{1}{\gamma _{e}\gamma _{n}\hbar ^{2}}\sum_{j}F_{j,r} \chi ^{\prime
}(j),  \label{K}
\end{equation}
where
\begin{equation}
\chi ^{\prime }(j)=-\frac{\mu _{B}^{2}}{\pi }\int d\varepsilon P
(\varepsilon ,T)\text{Tr }\text{Im}G\left( j,j,\varepsilon \right) ,
\label{x'}
\end{equation}

\begin{figure}[tb]
\includegraphics[width=6cm]{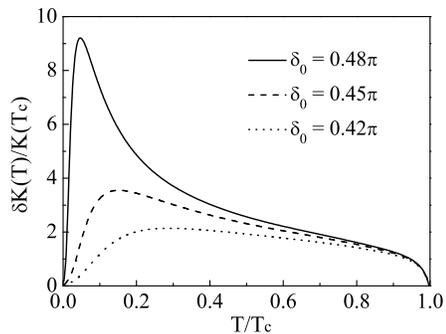} \vspace{-0.2cm}
\caption{The Knight shift versus temperature on one of the nearest neighboring sites of
impurity. $K(T_c)$ is the total Knight shift at $T_c$. $ A=-4B$ is taken and the other
parameters are the same as for Fig. \ref{fig1}.} \label{fig2}
\end{figure}

In the unitary limit, $(\Omega ^{\prime },\Omega ^{\prime \prime })\ll
\Delta $, the contribution from the resonance to $\chi ^{\prime }$ is
approximately given by
\begin{equation*}
\delta \chi ^{\prime }(j)\approx -\frac{\mu
_{B}^{2}\mathrm{Tr}\mathrm{Re}
G^{0}(j,0)\mathrm{Re}G^{0}(j,0)}{\pi }\int d\varepsilon
T_{11}^{\prime \prime }(\varepsilon) P(\varepsilon ,T).
\end{equation*}
As for $\chi ^{\prime \prime }$, the temperature dependence of $\delta \chi ^{\prime
}(j)$ is predominantly determined by the resonance pole; thus
\begin{equation*}
\delta \chi ^{\prime }(j)\sim P\left( \Omega ^{\prime },T\right) .
\end{equation*}
Again, in the intermediate-temperature regime $k_{B}T_{c}\gg k_{B}T\gg \Omega ^{\prime
}$, $P\left( \Omega ^{\prime },T\right) \sim 1/T$, from Eq. ( \ref{x'}) we then have
\begin{equation}
\delta K(r)\sim \frac{1}{T}.
\end{equation}
Thus the resonance state lead to a Curie-like term in the Knight shift. In the
low-temperature limit, $k_{B}T\ll \Omega ^{\prime }$, $P(\Omega ^{\prime },T)$ drops
exponentially with temperature. This suppresses the divergence of $K(T)$ in low
temperatures and $K(T)$ becomes zero at zero temperature. This is different from the
behavior of a free local moment. The overall temperature dependence of the impurity
contribution to the Knight shift $\delta K(r)$ on one of the nearest-neighboring sites of
the impurity is shown in Fig. \ref{fig2}. On other sites, for example, on one of the Y
sites closest to an impurity in YBCO, the impurity correction to $\delta K(r)$ is smaller
but the overall temperature dependence behaves similarly.

The above results indicate that the impurity contribution to the Knight shift follows the
Curie law in a broad temperature regime in the superconducting state. Thus the resonance
state induced by an impurity is equivalent to a local magnetic moment in the magnetic
measurement. This suggests that an effective moment corresponding to a resonance state
can be defined from the magnetic susceptibility by the following equation,
\begin{equation}
\frac{\mu _{eff}^{2}}{3k_{B}T}=\sum_{j}\delta \chi ^{\prime
}\left( j\right), \label{meff}
\end{equation}
where the summation runs over the four nearest neighbors of the impurity. This effective
moment can be used to characterize a nonmagnetic resonance state in the analysis of
Knight shift data.

\begin{figure}[tb]
\includegraphics[width=6cm]{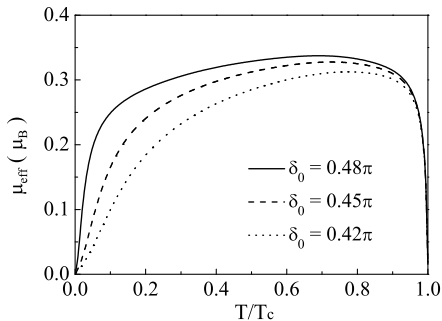} \vspace{-0.2cm}
\caption{The effective moment of a single nonmagnetic impurity
versus temperature.} \label{fig3}
\end{figure}

Figure \ref{fig3} shows the effective moment $\mu _{eff}$ corresponding to the resonance
state.
It should be emphasized that both $\delta \chi ^{\prime }$ and the effective moments
depend on the energy dispersion of normal electrons. The value of effective moment can be
altered if an energy dispersion other than that given by Norman \textit{et al.} is used.
\cite{Norman} In particular, $\delta \chi ^{\prime }$ is proportional to the normal state
density of states $N_{F}$. If $N_{F}$ is doubled, then $\mu _{eff}$ will increase by a
factor $\sqrt{2}$. Close to $T_{c}$ or at very low temperatures, the effective moment of
the resonance state becomes small and approaches zero in the zero-temperature limit. This
is different from a free magnetic moment and can in principle be used to separate the
contribution of the resonance state from that of a free magnetic moment.

If the temperature is not too low or too close to the critical temperature, the effective
moment is about 0.3$\mu _{B}$. This is close to the effective moment induced by Zn for
slightly overdoped YBCO deduced from the magnetic susceptibility data. \cite{Mendels}
However, it is much smaller than the effective moment induced by Zn for underdoped YBCO
or by Li for optimal doped YBCO. This indicates that the resonance state induced by a
strong nonmagnetic impurity has substantial contribution to the NMR spectra in the
high-$T_c$ superconducting state. This contribution should be taken into account in the
analysis of the NMR data. However, the effective moment of the resonance state is smaller
than the total moments induced by a nonmagnetic impurity in the underdoped regime. The
difference between the total moment determined from magnetic measurements and the
effective moment of the resonance state can be attributed to the contribution of local
spins induced by a nonmagnetic impurity.

In conclusion, we have studied the effect of nonmagnetic impurities on the NMR spectra of
high-$T_c$ superconductors. The resonance state near the Fermi surface induced by a
unitary impurity behaves effectively like a magnetic moment in the $d$-wave
superconducting state. It contributes a Curie-Weiss term to the NMR spin-lattice
relaxation rate as well as the Knight shift in the temperature regime $\Omega^\prime \ll
k_BT \ll k_BT_c$. The contribution of this induced resonance state can account for the
main feature of the NMR spectra in the superconducting state in overdoped high-$T_c$
materials. In the underdoped regime, the contribution of the resonance state to the NMR
spectra is also substantial, but the effective moment of the resonance state is smaller
than the total moments induced by a nonmagnetic impurity.

We thank L. Yu and G. M. Zhang for useful discussions. This work
was supported by the National Natural Science Foundation of China.

\end{document}